\documentclass[apj]{emulateapj} 
\usepackage{lscape} 
\usepackage{apjfonts}
\usepackage{latexsym}
\usepackage{graphicx,natbib}
\usepackage{hyperref}
\usepackage{amsmath} 
\usepackage{subfigure}
\usepackage{url}

\shorttitle{}
\shortauthors{A. Monachesi et al.}
\begin{document}

\title{GHOSTS I: A new faint very isolated dwarf galaxy at D =
  12$\pm2$ Mpc}

\author{Antonela Monachesi\altaffilmark{1}}\email{antonela@umich.edu}
\author{Eric F. Bell\altaffilmark{1}}\author{David
  J. Radburn-Smith\altaffilmark{2}} \author{Roelof S. de
  Jong\altaffilmark{3}}\author{Jeremy Bailin\altaffilmark{4}}
\author{Julianne J. Dalcanton\altaffilmark{2}}\author{Benne
  W. Holwerda\altaffilmark{5}}\author{H. Alyson
  Ford\altaffilmark{6}}\author{David
  Streich\altaffilmark{3}}\author{Marija
  Vlaji\'c\altaffilmark{3}}\author{Daniel B. Zucker \altaffilmark{7}}

  \altaffiltext{1}{Department of Astronomy, University of Michigan,
    830 Dennison Bldg., 500 Church St., Ann Arbor, MI 48109, USA}
  \altaffiltext{2}{Department of Astronomy, University of Washington,
    Seattle, WA 98195,USA} \altaffiltext{3}{Leibniz-Institut f\"ur
    Astrophysik Potsdam, D-14482 Potsdam,
    Germany}\altaffiltext{4}{Department of Physics \& Astronomy,
    University of Alabama, Box 870324, Tuscaloosa, AL 35487,
    USA}\altaffiltext{5}{European Space Agency Research Fellow
    (ESTEC), Keplerlaan 1, 2200 AG Noordwijk, The Netherlands}
  \altaffiltext{6}{National Radio Astronomy Observatory, P.O. Box 2,
    Green Bank, WV 24944, USA} \altaffiltext{7}{Department of Physics
    and Astronomy, E7A 317, Macquarie University, NSW 2109, Australia}
  \footnotetext[8]{Based on observations made with the NASA/ESA Hubble
    Space Telescope, obtained at the Space Telescope Science
    Institute, which is operated by the Association of Universities
    for Research in Astronomy, Inc., under NASA contract NAS 5-26555.}

\begin{abstract}
 We report the discovery of a new faint dwarf galaxy, GHOSTS I, using
 HST/ACS data from one of our GHOSTS (Galaxy Halos, Outer disks,
 Substructure, Thick disk, and Star clusters) fields. Its detected
 individual stars populate an approximately one magnitude range of its
 luminosity function (LF). Using synthetic color-magnitude diagrams
 (CMDs) to compare with the galaxy's CMD, we find that the colors and
 magnitudes of GHOSTS I's individual stars are most consistent with
 being young helium-burning and asymptotic giant branch stars at a
 distance of $\sim 12\pm2$ Mpc.  Morphologically, GHOSTS I appears to
 be actively forming stars, so we tentatively classify it as a dwarf
 irregular (dIrr) galaxy, although future HST observations deep enough
 to resolve a larger magnitude range in its LF are required to make a
 more secure classification. GHOSTS I's absolute magnitude is $M_{V}
 \sim -9.85^{+0.40}_{-0.33}$, making it one of the least luminous dIrr
 galaxies known, and its metallicity is lower than $[\mathrm{Fe/H}] =
 -1.5$ dex. The half-light radius of GHOSTS I is $226\pm38$ pc and its
 ellipticity is $0.47 \pm 0.07$, similar to Milky Way and M31 dwarf
 satellites at comparable luminosity. There are no luminous massive
 galaxies or galaxy clusters within $\sim 4$ Mpc from GHOSTS I that
 could be considered as its host, making it a very isolated dwarf
 galaxy in the Local Universe.
\end{abstract}


\maketitle

\section{Introduction}\label{sec:intro}

Despite the rapid rate of discovery of new Local Group dwarf galaxies
\citep[e.g.,][see \citealt{McConnachie12} for a review]{Willman05,
  Belokurov07, Martin09, Slater11, Richardson11}, the number of
observed dwarf galaxies is far lower than the number of dark matter
halos expected around massive galaxies in a cosmological context
\citep[e.g.,][]{Moore99, Klypin99}. It is widely thought that the most
reasonable interpretation of this difference is that most very low
mass halos do not host detectable stellar content for a number of
possible reasons, including effective feedback from star formation or
inefficient gas cooling and accretion \citep[e.g.,][]{DekelSilk86,
  Bullock00, Benson02, Koposov09}. Yet, many modeling uncertainties
remain: for example, dramatic variations in galaxy formation
efficiency at low mass may be required \citep{Boylan-Kolchin12},
feedback-driven changes in the central density profile of dwarf
galaxies may dramatically reduce the number of surviving Local Group
dwarf galaxies \citep{Brooks13}, and at any rate there is a
significant degree of halo-to-halo scatter expected in dwarf galaxy
luminosity functions \citep{Gomez12}.

One of the interesting open issues is how dwarf galaxies are affected
by being close to giant (e.g., Milky Way mass) galaxies. The tidal and
ram pressure effects of such a halo are considerable and likely
transform many star-forming dwarf irregular [dIrr] galaxies into
non-star-forming dwarf spheroidal [dSph] galaxies
\citep[e.g.,][]{Mayer01, Mayer06}. Indeed, it appears quite plausible
that even a single pass by a massive galaxy is sufficient to shut off
star formation in a dwarf galaxy, leaving a dwarf spheroidal remnant
(\citealt{Slater_bell13}; see also \citealt{Teyssier12} and
\citealt{Kazantzidis13}); such processes can create dSph galaxies out
to more than 1 Mpc from the giant galaxy.

Tidal forces strongly affect dwarf galaxies, and those experiencing
large tidal forces will get disrupted \citep[e.g.,][]{Brooks13},
leaving an extended stellar halo around massive galaxies
\citep{Bullock01, BJ05, Bell08,Cooper10}. Yet, current surveys are
sensitive to ultra-faint $M_V\gtrsim -6.5$ dwarfs only within 300
  kpc of the Milky Way (SDSS; \citealt{Walsh09}) and M31 (PANDAS;
\citealt{Richardson11}). Only brighter systems can be seen at larger
distances from the Milky Way \citep[e.g.,][]{Irwin07}. Consequently,
it is possible that our current dwarf galaxy luminosity functions
\citep[e.g.,][]{Koposov08, Walsh09} are incorrect, particularly so at
the faint end.

 In this context, it is important to make progress towards the
 long-term goal of characterizing the number and properties of dwarf
 galaxies very far from giant galaxies. The M81 Group contains
 $\sim40$ dwarf galaxies brighter than $M_I \sim -9$
 \citep{Chiboucas09, Chiboucas13}; they all lie within $1.2$ Mpc. The
 Cen A group of galaxies contains $\sim$ 60 dwarfs, with its faintest
 galaxy having a total magnitude of $M_B\sim -10$
 \citep{Karachentsev02}. The two farthest dwarfs from this group are
 at $\sim 2$ Mpc from Cen A and have a total magnitude of $M_B\sim
 -12$ \citep{Crnojevic12}. Furthermore, most distant dwarf galaxies in
 both groups, as well as in the Local Group, appear to be star-forming
 dIrrs, although with some important exceptions (see
 \citealt{Pasquali05} for a discussion of APPLES 1, a field dwarf
 spheroidal galaxy at D = 9 Mpc not associated with any major cluster
 of galaxies and \citealt{Makarov12} for a discussion of KKR 25, a
 dwarf spheroidal galaxy 1.9 Mpc from the Milky Way). In this work, we
 report on the discovery of a very isolated faint dwarf galaxy (which
 appears to be a dIrr), projected close to M81 on the sky but at
   least $\sim 6$ Mpc more distant than M81 and with a total absolute
   magnitude of $M_V = -9.85^{+0.40}_{-0.33}$.

\section{Observations and discovery}\label{sec:observations}

The GHOSTS\footnote{GHOSTS=Galaxy Halos, Outer disks, Substructure,
  Thick disk, and Star clusters} survey (PI = R. de Jong, HST programs
= 10523, 10889, 11613, and 12213) consists of \emph{HST} ACS/WFC and
WFC3/UVIS images of various fields in the outskirts of nearby disk
galaxies. One of its main goals is to resolve the stars in the halos
and thick disks of the sampled galaxies, giving insight into their
masses, stellar populations, structures, and degree of
substructure. The images are reduced homogeneously through the GHOSTS
pipeline \citep[][hereafter R-S11]{RS11}. PSF stellar photometry of
each image was performed using \verb+DOLPHOT+ \citep{Dolphin00} and
various diagnostic parameters were used to discriminate spurious
detections (mostly contamination from unresolved galaxies) from the
actual stars. Several selection criteria to discriminate unresolved
galaxies from stars were optimized using deep archival high-redshift
HST/ACS fields. These were applied to the raw photometric output,
which removed $\sim$95\% of the contaminants. We refer the reader to
R-S11 for full details of the GHOSTS data pipeline and the photometric
culls. The final catalog of stars is largely free of background
contaminants.

\begin{figure}\centering
\includegraphics[width=88mm,clip]{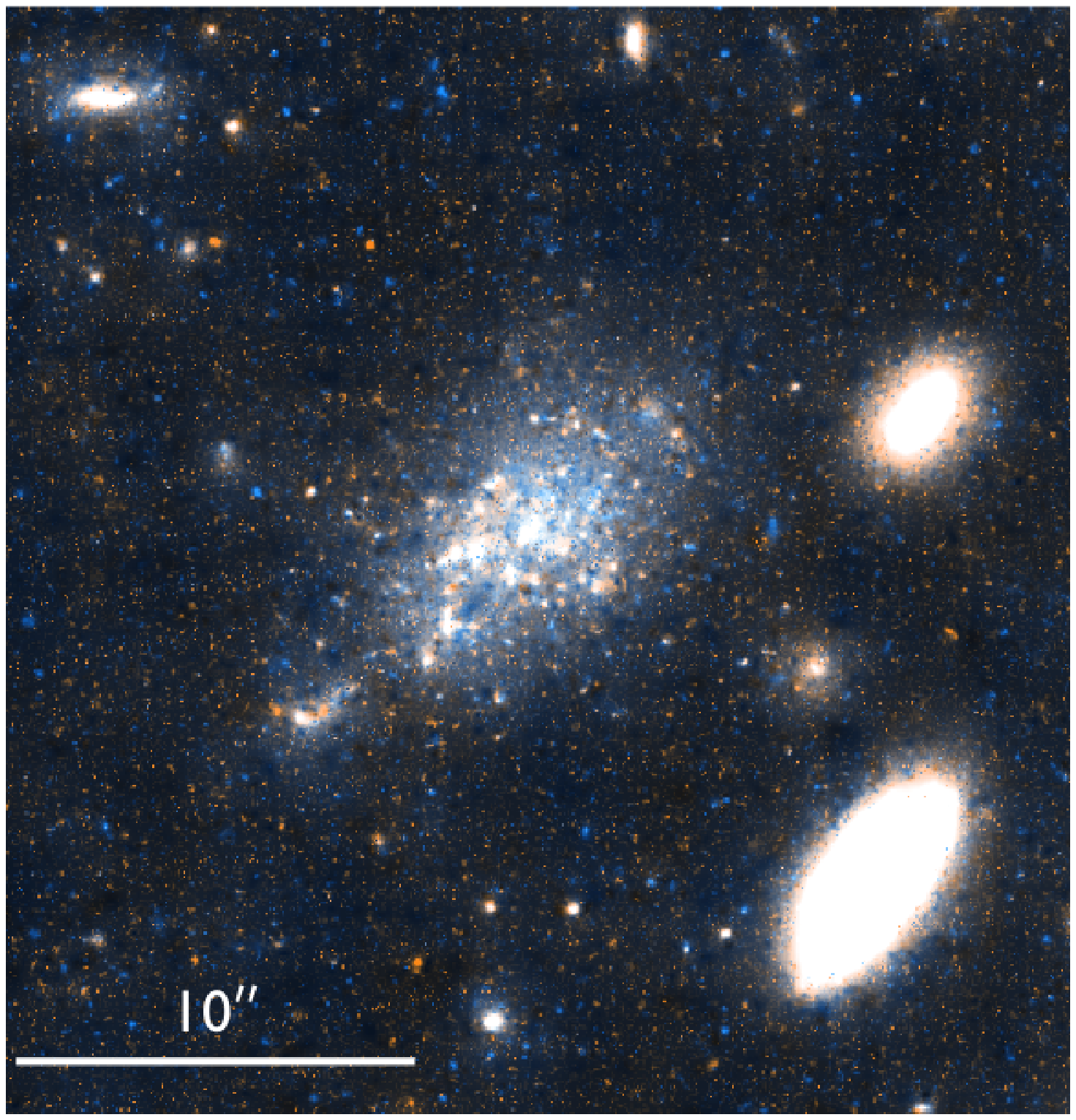}
\includegraphics[width=88mm,clip]{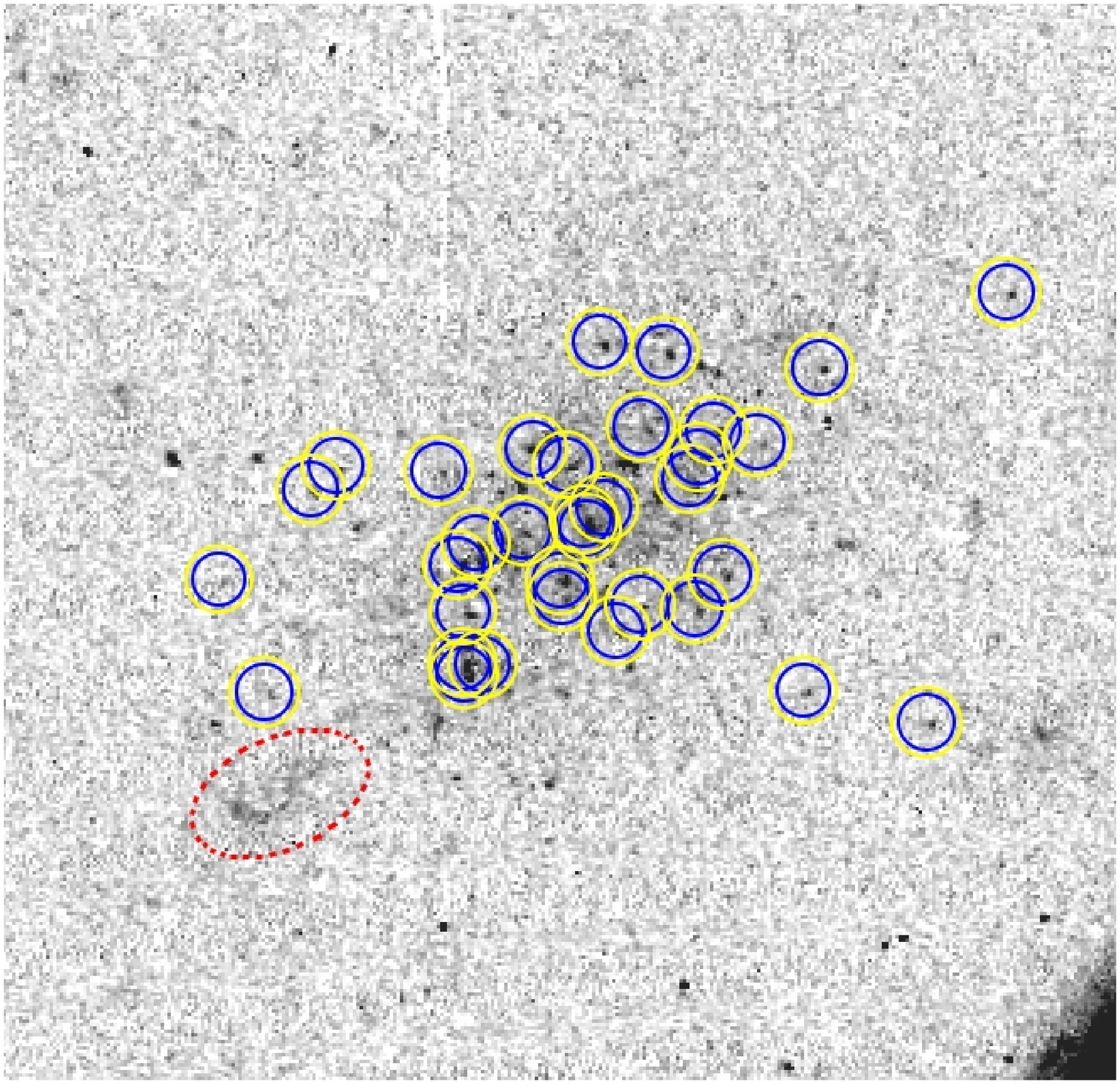}
\caption{HST image of GHOSTS I. Top panel: Color image of GHOSTS I, in
  one of the GHOSTS fields located at a projected galactocentric
  distance of 32 kpc from M81, at M81's distance. The image size is
  $\sim 24.5\arcsec \times 24.5\arcsec$. Bottom panel: Zoom in of the
  $F606W$-band image of GHOSTS I, covering an area of $\sim 15
  \arcsec.8\times 15\arcsec.8$. Circles indicate the individual stars
  in the region of the dwarf detected using the GHOSTS pipeline and
  the red dashed ellipse indicates a possible star forming
  region. North is up and east is to the left. Note that some of
    the 3 circles in the lower left may be deblends of one source.}
\label{fig:dwarf}
\end{figure}

Visual inspection of the drizzled GHOSTS images revealed a dwarf
galaxy candidate, partially resolved into individual stars, in one of
the HST/ACS fields located in the halo of M81 (HST program 11613, PI =
de Jong, Field 16 in \citealt{Monachesi13}). The drizzled HST/ACS
image of each $F606W$ and $F814W$ filter is produced using the STScI
Multidrizzle
routine\footnote{\url{http://stsdas.stsci.edu/multidrizzle/}} after
combining only two exposures per filter, and therefore detection and
rejection of cosmic rays is challenging. Indeed, there were some
obvious cosmic rays left in the images when the default values for the
Multidrizzle parameters were used. Since contamination by cosmic rays
hamper the detection of individual stars, we performed a more careful
cosmic ray rejection in the region around the dwarf by using the
pixel-based charge transfer efficiency corrected images and optimizing
the parameter values, especially those for creating the median image
and the cosmic ray masks that are used by Multidrizzle when combining
the images. The new HST/ACS combined images, both in $F606W$ and
$F814W$, clearly show the dwarf galaxy, GHOSTS I, and the 34
individual stars that can now be resolved. We show in
Figure~\ref{fig:dwarf} a color HST image of GHOSTS I, in the top
panel, and a zoomed-in $F606W$-band HST image with its individual
stars indicated by circles in the bottom panel. To estimate the
  overdensity of stars in the dwarf region, we counted the stars
  inside 150 randomly--selected areas in the field, covering the same
  area as the dwarf region \footnote{Regions near the massive
    globular cluster discovered in this field \citep{Jang12} were
    discarded. We find that the mean number of stars inside the
  sample of 150 regions is 5 with a standard deviation of $\sigma =
  2.25$. Assuming a Poisson distribution in the field of n=5 stars per
  area, the probability of finding 34 stars within the same area is
  $1.3\times10^{-17}$.}

\begin{figure*}\centering
 \includegraphics[width=180mm,clip]{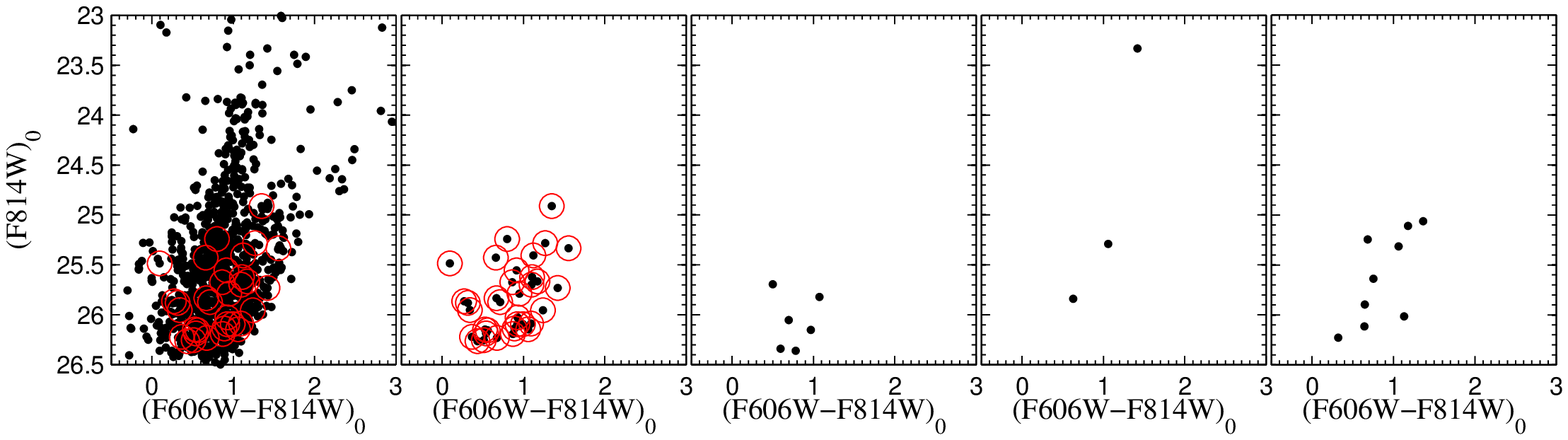}
\caption{First panel: CMD of stars in M81's halo field with the
  probable dwarf member stars highlighted with red circles. Second
  panel: CMD of stars within $10 \arcsec\times 5\arcsec$ ($200\times 100$
  pixels$^2$) centered at the dwarf. Third to fifth panels: CMDs of
  stars within $10 \arcsec\times 5\arcsec$ ($200\times 100$ pixels$^2$)
  around three randomly chosen field locations. There is a clear
  overdensity of stars where the dwarf is located. More precisely,
  there are $34\pm5.8$ stars within the dwarf region, where the
  uncertainty is due to Poisson error, whereas the mean number of
  stars within a same area of 150 randomly--selected regions in the
  field is $5\pm2.25$.}
\label{fig:cmds}
\end{figure*}

Figure~\ref{fig:cmds} shows, in the left panel, the color-magnitude
diagrams (CMD) of all objects classified as stars by the GHOSTS
pipeline for Field 16. Red circles represent the stars spatially
coincident with the dwarf. The magnitudes have been corrected for
Galactic extinction using the corrected extinction ratios
$A_{\lambda}/E(B-V)$ of 2.47 and 1.53 for $F606W$ and $F814W$,
respectively, that are to be used with the E(B-V) values from
\citet{Schlegel98} dust maps and an $R_V=3.1$ extinction law
\citep{Schlafly11}. The second to fifth panels, from left to
  right, show the stars detected inside the dwarf region, for an area
  of $10 \arcsec\times 5\arcsec$ ($200\times 100$ pixels$^2$) centered
  at GHOSTS I, and inside three random places in the image covering
  the same area, respectively. The overdensity of stars around the
dwarf galaxy is evident.

After discovering GHOSTS I, we checked the Sloan Digital Sky Survey
(SDSS) data to see if it was detected by the survey. We found that
GHOSTS I was photometrically identified by SDSS and classified as
  a galaxy at RA= 09:53:13.75, DEC= $+$69:30:01.58. There is only
information about its integrated photometry and no studies about this
galaxy have been done so far.

\section{Properties of GHOSTS I}\label{properties}

To measure the apparent magnitude of GHOSTS I we perform aperture
photometry around the galaxy using the IMEXAMINE/IRAF task. We use an
aperture radius of 100 pixels, equivalent to $5{\arcsec}$. We estimate
the background contamination by averaging the luminosity of four
randomly selected regions across the field, for which we also
performed aperture photometry using the same aperture radius. The
apparent Galactic extinction--corrected magnitudes of GHOSTS I in
Vega-mag $F606W$ and $F814W$ filters are $20.36\pm0.2$ and
$19.73\pm0.15$, respectively. These magnitudes were converted into $V=
20.60\pm0.28$ and $I=19.54\pm0.15$ using the transformations between
the HST/ACS and BVRI photometric systems by \citet{Sirianni05}. Its
integrated color is $V-I = 1.06 \pm 0.31$. The SDSS integrated colors
for this galaxy are $g-r=0.45\pm0.11$ and $g-i=0.37\pm0.15$.

\begin{figure}\centering
 \includegraphics[width=85mm,clip]{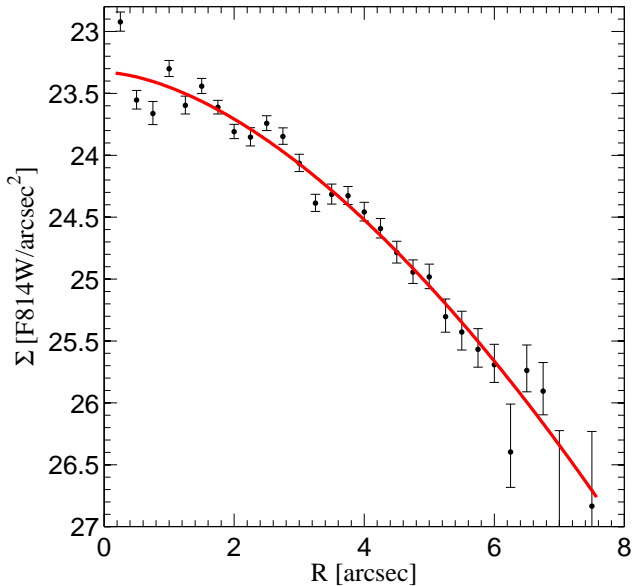}
\caption{$F814W$--surface brightness profile of GHOSTS I, as obtained
  by the isophotal fitting. The red curve indicates a Sersic profile
  fit. The half-light radius and Sersic index obtained from this fit
  are $r_h= 3\arcsec.9 \pm 0\arcsec.3$ (or $149\pm 11$ pc at the dwarf
  distance), and $n=0.6 \pm 0.17$, respectively.}
\label{fig:profiles}
\end{figure}

To obtain the galaxy's ellipticity, we fit elliptical isophotes to the
dwarf's image using the IRAF task ELLIPSE. The galaxy's ellipticity
and position angle were determined from the outer isophotes in the
F814W-band image. To calculate the dwarf's half-light radius, we fit a
Sersic profile function to the surface brightness profiles in both
filters, using the isophotal results (see
Figure~\ref{fig:profiles}). The dwarf's properties, as well as its sky
location, are listed in Table~\ref{table:properties}.

\begin{deluxetable}{lc}
  \tabletypesize{\scriptsize} \tablecaption{ Properties of GHOSTS
    I\label{table:properties}} \tablewidth{150pt} \centering
  \tablehead{\colhead{Parameter}& \colhead{Value}} \startdata $\alpha$
  (J2000) &
  $09^{\textrm{h}}53^{\textrm{m}}13^{\textrm{s}}.7$\\ $\delta$ (J2000)
  & $69^{\circ}30'02''.63$\\ E(B-V)\tablenotemark{a} & 0.091 \\ D
  [Mpc] & $12 \pm
  2$\\ $M_V$ & $-9.85^{+0.40}_{-0.33}$ \\ Ellipticity
  & $0.47 \pm 0.07$ \\ Position angle (N to E) &
  $-39^{\circ}.43\pm10^{\circ}$ \\ $r_h$ [$\arcsec$] & $3.9 \pm 0.3$
  \\ $r_h$ [pc] & $226\pm 38$
  \\ $(m-M)_0$ & $30.40^{+0.33}_{-0.40}$ \enddata
  \tablenotetext{a} {\citet{Schlegel98}.} 
\end{deluxetable} 

\section{The distance of GHOSTS I} \label{distance}

As shown in Fig~\ref{fig:cmds}, we only detect approximately a one
magnitude range of GHOSTS I's luminosity function (LF). Since we know
neither the distance nor the star formation history (SFH) of the
galaxy, the detected stars could be in principle consistent with
different stellar evolutionary phases seen at various distances. In
this section, we use model CMDs to evaluate what kinds of stars
  could be populating the observed LF. The considered models are
based on different assumptions of the galaxy's SFH and they allow us
to constrain the dwarf's distance and therefore its properties, such
as its total absolute magnitude and half-light radius.

We use a set of theoretical CMDs with different stellar populations
(i.e. different SFHs) generated using the IAC-STAR code
\citep{Aparicio_gallart04}. Each model CMD covers different ages
  and metallicities. The total age and metallicity ranges used are 0
  to 12 Gyr and Z= 0.0001 to 0.0019, respectively. For each model CMD,
  we assume various distances in the range of 4 to 18 Mpc. Given a
  distance, the absolute magnitudes of the model stars are transformed
  into apparent magnitudes. We then populate each model CMD with
  randomly extracted mock stars until the cumulative luminosity
  matches the observed luminosity of GHOSTS I. In other words, we
  generate model CMDs at different distances that have the same total
  luminosity as GHOSTS I.

To make a fair comparison between the properties of GHOSTS I's stars
and those from the model CMDs, we simulate the observational effects
(incompleteness and photometric errors) on the extracted mock stars
using the results obtained from the extensive artificial star tests
(ASTs) as follows \citep[see also][]{Monachesi12, Monachesi13}. Each
star in the model CMD is assigned a magnitude and color correction
from the AST results. This correction is the difference between the
injected and recovered magnitudes of a randomly selected artificial
star of similar magnitude and color to the model star. If the randomly
selected artificial star is lost (due to crowding or magnitude
detection limit), the model star is also considered "lost". Therefore,
after the results from the ASTs are simulated, each model CMD is
carefully corrected for completeness and photometric errors as a
function of both magnitude and color.

 As a last step before comparing the observed CMD against models, we
 need to take into account the foreground contamination of M81 field
 stars. We have shown in the previous section that M81 field stars
 overlap in color and magnitude with the dwarf galaxy stars. Using the
 150 random field CMDs (see section~\ref{sec:observations}), we have
 generated an average foreground CMD by calculating the mean number of
 field stars as a function of magnitudes. The magnitudes and colors
 assigned to those stars per magnitude bin were calculated using the
 mean and standard deviation values from all the field stars in that
 particular magnitude bin.  The foreground CMD was added to each model
 CMD.

Finally, the resulting (model $+$ foreground) CMD is quantitatively
compared with the CMD of GHOSTS I to assess the viability of each
model and each distance. Specifically, we compare the number of stars
as a function of magnitude. For the cases where the number of stars
coincides, within Poisson uncertainties, their color distributions are
compared.  Figure~\ref{fig:distancelfs} shows the comparison between
the LF of GHOSTS I, indicated with a red line, and some of the models
(from top to bottom) at different distances (from left to right),
indicated with black dashed lines. The errorbars indicate Poisson
uncertainty. For the cases where both model and observed LFs agree
(letters A, B, C, and D in Figure~\ref{fig:distancelfs}) we compare
their color distributions, shown in Figure~\ref{fig:distancecolors}
with the same corresponding letter, and make the final decision as to
whether the model and specific distance are
acceptable. Figure~\ref{fig:distancecmds} shows some examples of the
model (black dots) and M81 foreground stars (gray dots) CMDs, which
are to be compared with the GHOSTS I CMD, shown in each panel as red
circles.

From these quantitative comparisons, we find that the most likely
distance of GHOSTS I is $12\pm2$ Mpc and that the stars that we
observe are very young (0--200 Myr) blue and red helium-burning stars
(BHeB and RHeB) as well as young AGB stars.  In what follows, we
describe and discuss some of the relevant considered scenarios:

\begin{figure*}\centering
 \includegraphics[width=180mm,clip]{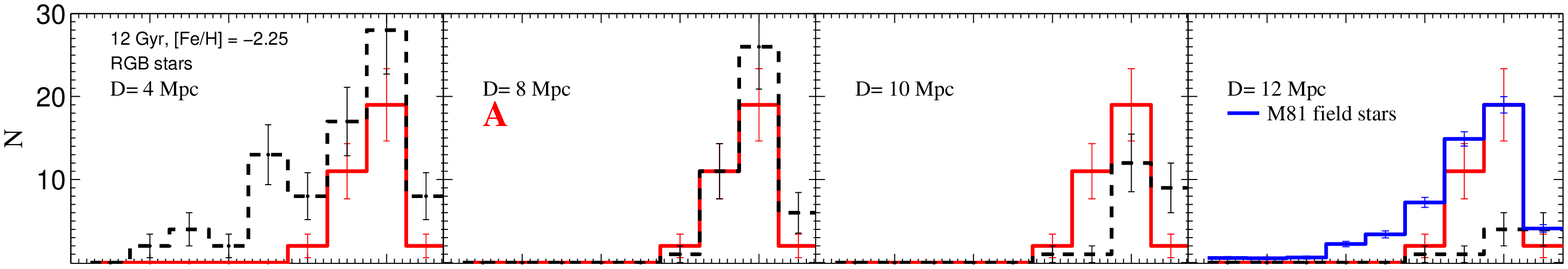}
\includegraphics[width=180mm,clip]{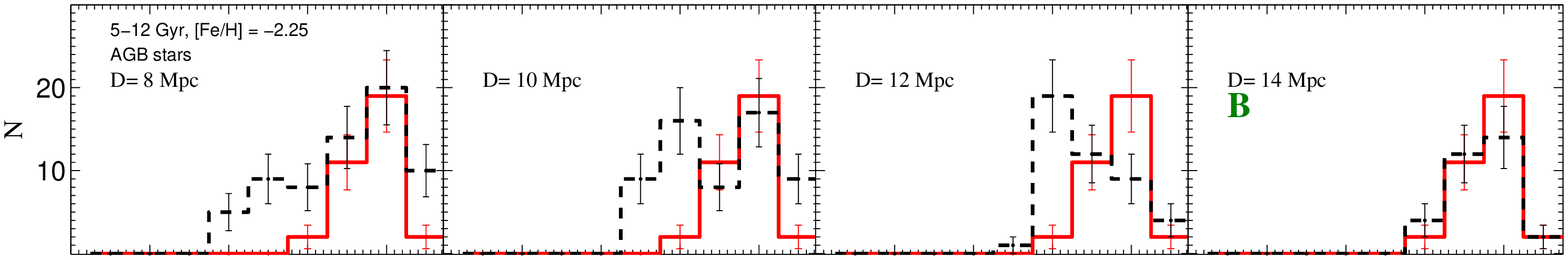}
\includegraphics[width=180mm,clip]{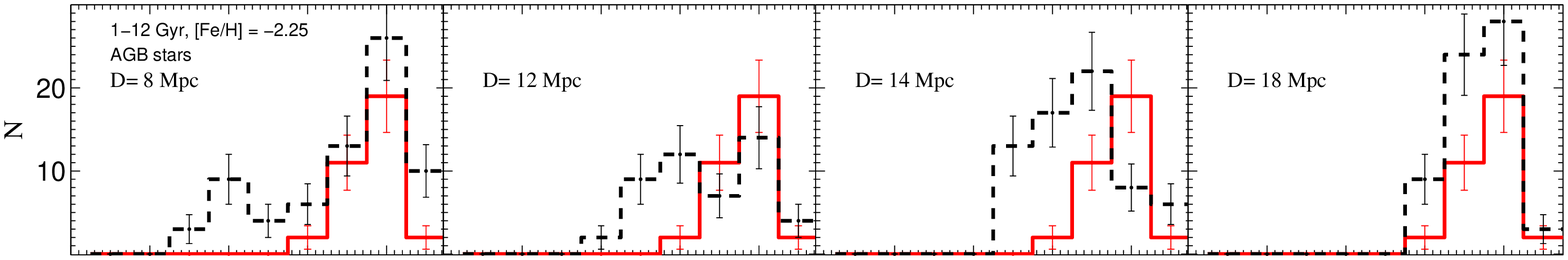}
\includegraphics[width=180mm,clip]{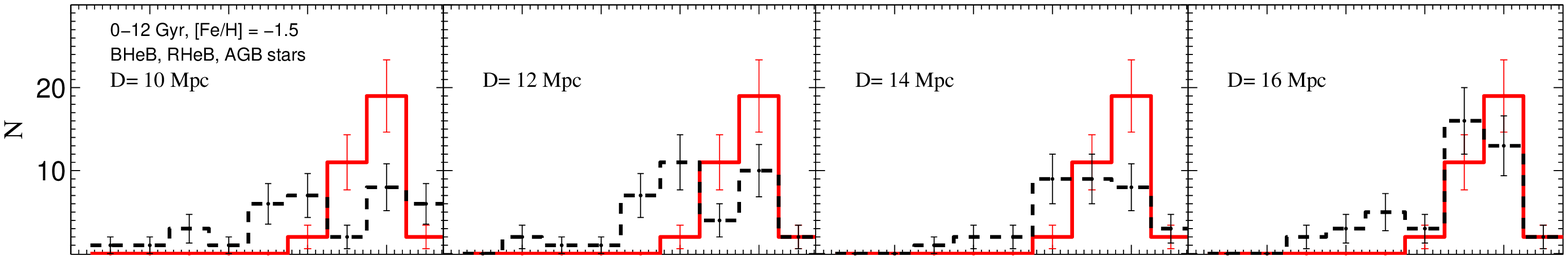}
\includegraphics[width=180mm,clip]{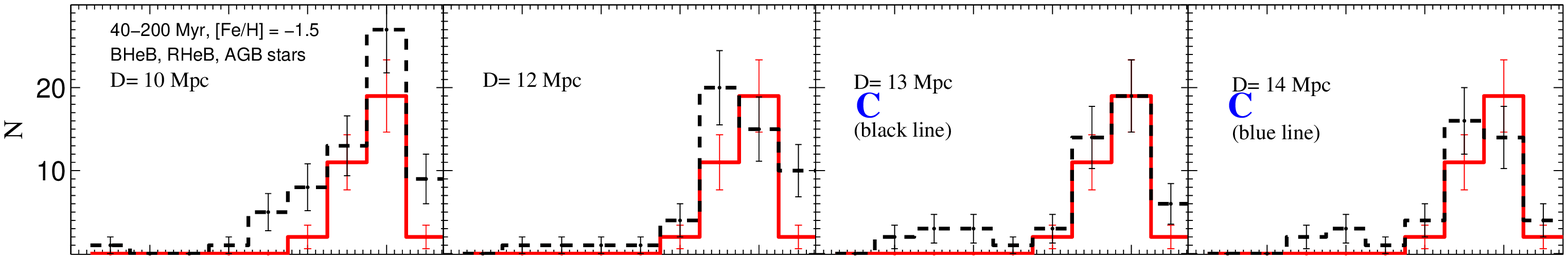}
\includegraphics[width=180mm,clip]{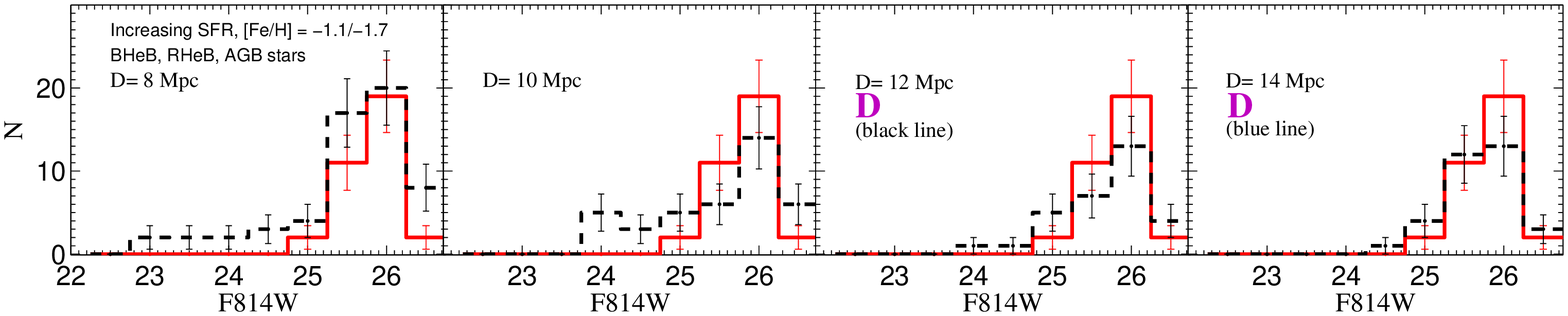}
\caption{Comparison of number of stars as a function of
    magnitudes between GHOSTS I and different model CMDs that have
    GHOSTS I's total apparent magnitude at various distances. Red
    lines in all panels show the LF of GHOSTS I. Each row indicates a
    different model CMD, which has been corrected for observational
    effects and has added an average foreground contamination from M81
    field stars.  From left to right of each panel we show the
    resulting model LFs, as black dashed lines, assuming different
    distances. Errorbars indicate Poisson uncertainties. The panels
    labeled as $A$, $B$, $C$, and $D$ are the ones that match the LF
    of GHOSTS I and their color distributions are shown in
    Figure~\ref{fig:distancecolors} with the same labels. The blue
    line in the rightmost panel of the first row shows the LF of M81
    field stars, scaled to the LF of GHOSTS I. The shape difference
    between both LFs indicate that GHOSTS I has different stellar
    populations than M81's halo and that it is not at M81's
    distance. See the text for more details.}
\label{fig:distancelfs}
\end{figure*}

\begin{itemize}

\item {\it Faint RGB stars at M81's distance (D $\sim$ 3.7 Mpc)}: 
  A first straightforward hypothesis to analyze is that GHOSTS I is a
  satellite of M81. The dwarf galaxy's resolved stars could be red
  giant branch (RGB) stars at the distance of M81, since their apparent
  magnitudes and colors seem consistent with those of faint RGB stars
  in the M81's field CMD (see leftmost panel in
  Fig~\ref{fig:cmds}). From the LFs comparison procedure outlined
  above, we find that \emph{none} of the models at 4 Mpc, no matter
  what stellar populations we use, agrees with the LF of GHOSTS I. If
  GHOSTS I is at M81's distance, we should have been able to
  statistically observe at least 7 RGB stars in the dwarf brighter
  than $F814W \sim 25$ (see, e.g., left panel in the first row of
  Figure~\ref{fig:distancelfs}) whereas we observe none. For a mean of
  7, a Poisson distribution gives a probability of $9\times10^{-4}$ of
  observing 0 stars. In addition, a LF of M81 field stars, normalized
  to GHOSTS I's LF can be seen in the rightmost panel of the first row
  in Figure~\ref{fig:distancelfs} as blue line. The different shape in
  their luminosity functions indicates that GHOSTS I does not have a
  similar stellar population, and thus it is not a satellite, of
  M81. We therefore rule out the possibility that the dwarf is at
  M81's distance.

\item {\it Old metal poor TRGB stars}: If we consider the case in
  which the detected stars are old (12 Gyr) metal poor
  ($[\mathrm{Fe/H}]=-2.25$ dex) RGB stars, the brighter stars that we
  observe would be the tip of its red giant branch (TRGB) from which
  we can estimate the distance of the dwarf galaxy. The absolute $I$
  magnitude of the TRGB is rather constant ($M_{I} \simeq -4 $) for
  populations older than $\sim 3$ Gyr with metallicities lower than
  $[\mathrm{Fe/H}]\sim -0.7$ \citep{Bellazzini01}. Thus, this
  evolutionary feature is commonly used for dwarf galaxies to
  determine their distances. We find that the number of stars that we
  observe in GHOSTS I as a function of magnitude is consistent with a
  model of such population at D = 8 Mpc (see second from left to right
  panel from the top panel of Figure~\ref{fig:distancelfs}). However,
  the color distribution of these stars do not match the one of GHOSTS
  I (see first panels of Figures~\ref{fig:distancecolors} and
  ~\ref{fig:distancecmds}). Ages between 10 and 13.5 Gyr show similar
  results; however higher metallicity CMDs are even redder when
  compared with the color distribution.

\item {\it Tip of the AGB stars}: We consider the possibility that the
  observed stars are bright asymptotic giant branch (AGB) stars with
  ages between 1 and 7 Gyr. The AGB also shows a defined tip in the
  LF, as seen for example in F8D1 \citep{Dalcanton09}. As previously
  explained GHOSTS I appears to be a star forming galaxy. Thus, a
  bright AGB may very well be present in its CMD. Model CMDs generated
  with a constant star formation rate (SFR) from 12 to 1--7 Gyr ago
  contain bright AGB stars (brighter than the TRGB) of ages between
  1--7 Gyr. Some of these models, e.g., the one with a SFR from 12 to
  5 Gyr produce a LF that matches that of GHOSTS I (see second row of
  Figure~\ref{fig:distancelfs}). However, the mock AGB stars are too
  red when compared with the observed stars, regardless of the chosen
  metallicity (see second panel of
  Figure~\ref{fig:distancecolors}). Figures~\ref{fig:distancelfs},
  ~\ref{fig:distancecolors}, and ~\ref{fig:distancecmds} show the case
  in which the most metal poor population available (i.e. having the
  bluest colors) is considered. We can see from the second panel of
  figure~\ref{fig:distancecmds} that the mock bright AGB stars are
  significantly redder than the data.

\item {\it Young BHeB and RHeB stars}: Another possibility is that the
  stars that we observe are very young (0--200 Myr) blue and red
  helium-burning stars (BHeB and RHeB). In particular the RHeB
  sequence shows well defined tips in the LF, as seen in the dwarf
  galaxies Holmberg IX and Sextans A
  \citep{Dalcanton09}. Morphologically, GHOSTS I looks very clumpy,
  i.e. not as smooth as a dSph would be. This suggests that GHOSTS I
  could be actively forming stars and therefore the presence of HeB
  stars would not be unexpected. We analyzed different models where
  these kinds of stars are present, e.g. constant SFR from 0 to 12 Gyr
  ago (see forth row of Figure~\ref{fig:distancelfs}), old stars plus
  a burst of star formation from 200 to 40 Myr ago (fifth row of
  Figure~\ref{fig:distancelfs}), and an increased SFR during the last
  5 Gyr (last row of Figure~\ref{fig:distancelfs}), for metallicities
  between $[\mathrm{Fe/H}]=-1.5$ and $[\mathrm{Fe/H}]=-2.25$. As shown
  in Figures~\ref{fig:distancelfs}, \ref{fig:distancecolors}, and
  ~\ref{fig:distancecmds}, the last two models match nicely both the
  LF and color distribution of GHOSTS I's stars at distances of $12\pm
  2$ Mpc. The model with continuous SFR from lookback times of 12 to 0
  Gyr did not produce LFs that match the LF of GHOSTS I's stars.

\end{itemize}

We conclude from this quantitative analysis that the most favorable
scenario is the one in which GHOSTS I is a star forming galaxy at D
$=12\pm2$ Mpc. Follow up observations of GHOSTS I, in particular
deeper HST observations, would help to better constrain the distance
to GHOSTS I as well as its stellar population to confirm its nature.

\begin{figure*}\centering
\includegraphics[width=180mm,clip]{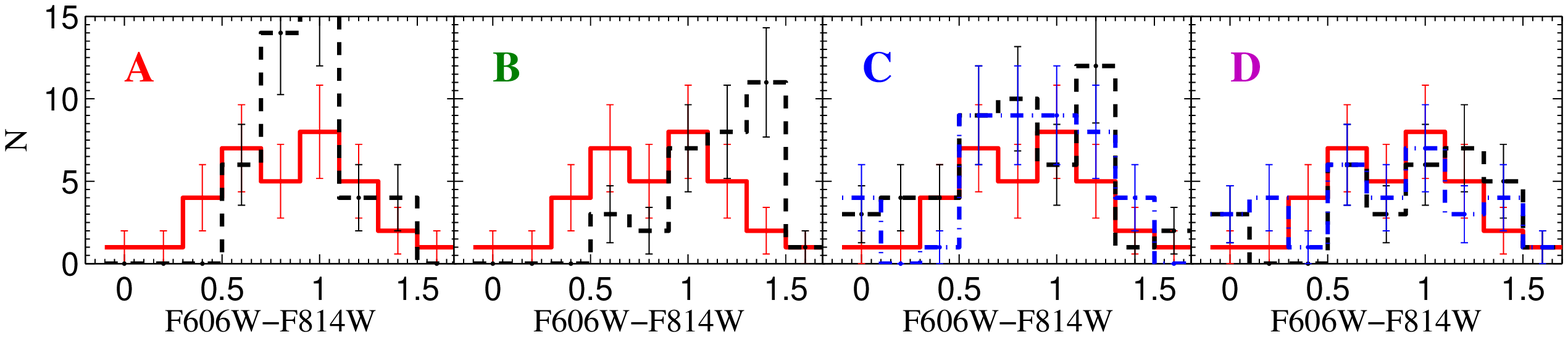}
\caption{Comparing the color distributions of model CMDs with
    GHOSTS I. The solid red line in each panel indicates the color
    distribution of GHOSTS I whereas the dashed lines show the color
    distributions of model CMDs, after the observational effect as
    well as the foreground M81 field stars contamination were taken
    into account. All these models have LFs that match the LF of
    GHOSTS I. The labels in each panel correspond to the labels in
    Figure~\ref{fig:distancelfs}, i.e. the model in panel $A$ is an
    old metal poor stellar population at 8 Mpc, $B$ corresponds to
    5---8 Gyr old AGB stars at 14 Mpc, $C$ corresponds to a SFH with a
    burst of star formation between 40 to 200 Myr at 13 and 14 Mpc,
    and $D$ shows the color distribution for a galaxy at 12 and 14 Mpc
    with an increased SFR during the last 5 Gyr. Clearly, the last two
    panels match better the color distribution of GHOSTS I.}
\label{fig:distancecolors}
\end{figure*}

\begin{figure*}\centering
 \includegraphics[width=180mm,clip]{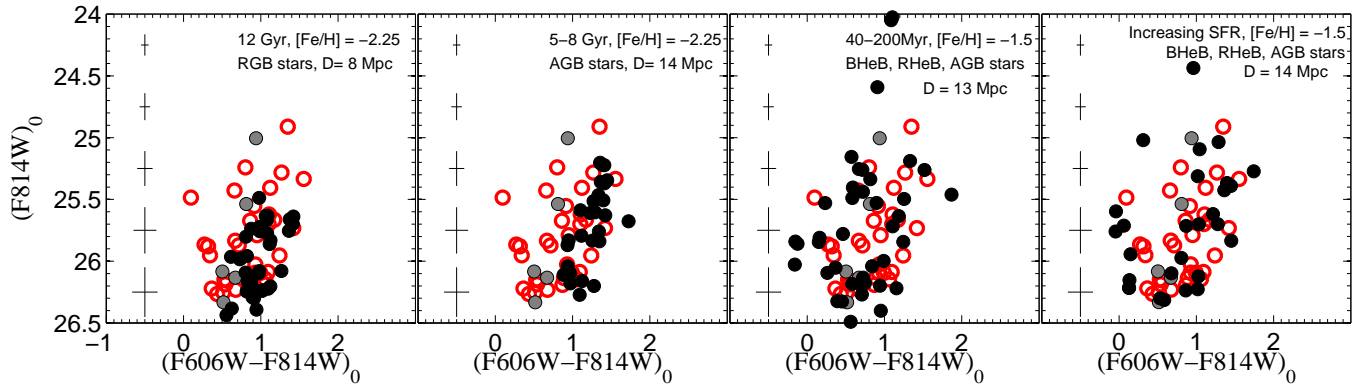}
\caption{Exploring possible scenarios about what kinds of stars
  populate the one magnitude range of CMD detected in GHOSTS I.
    In each panel, we show as red circles the stars at the dwarf's
    location, as black dots the mock stars after the observational
    effects were simulated for a galaxy having GHOSTS I's total
    apparent magnitude, and as gray dots the average foreground M81
    field stars. The errorbars indicate the photometric errors
    obtained from the ASTs and refer only to color $(F606W - F814W) =
    1$. Leftmost panel: The model CMD represents an old, metal poor
    population at 8 Mpc. The number of stars in the dwarf as well as
    their magnitudes are consistent with being the brightest portion
    of the RGB; however their colors distribution differ. Second
    panel: The brightest stars of this model CMD are AGB stars having
    ages between 5--8 Gyr at 14 Mpc. We can see that, even though
    their numbers and magnitudes agree with the stars in the dwarf,
    their colors are too red. Third panel: Very young HeB and AGB
    stars populate the one magnitude bright portion of the model CMD
    at a distance of 13 Mpc. The dwarf's stars are consistent with
    being such population. Rightmost panel: The resulting model CMD
    for a galaxy at D = 14 Mpc with increased SFR during the last 5
    Gyr. GHOSTS I's stars are also consistent with being drawn from
    such a population. See the text for more details.}
\label{fig:distancecmds}
\end{figure*}

\section{Discussion and conclusions}\label{conclu}

\begin{figure*}\centering
 \includegraphics[width=90mm,clip]{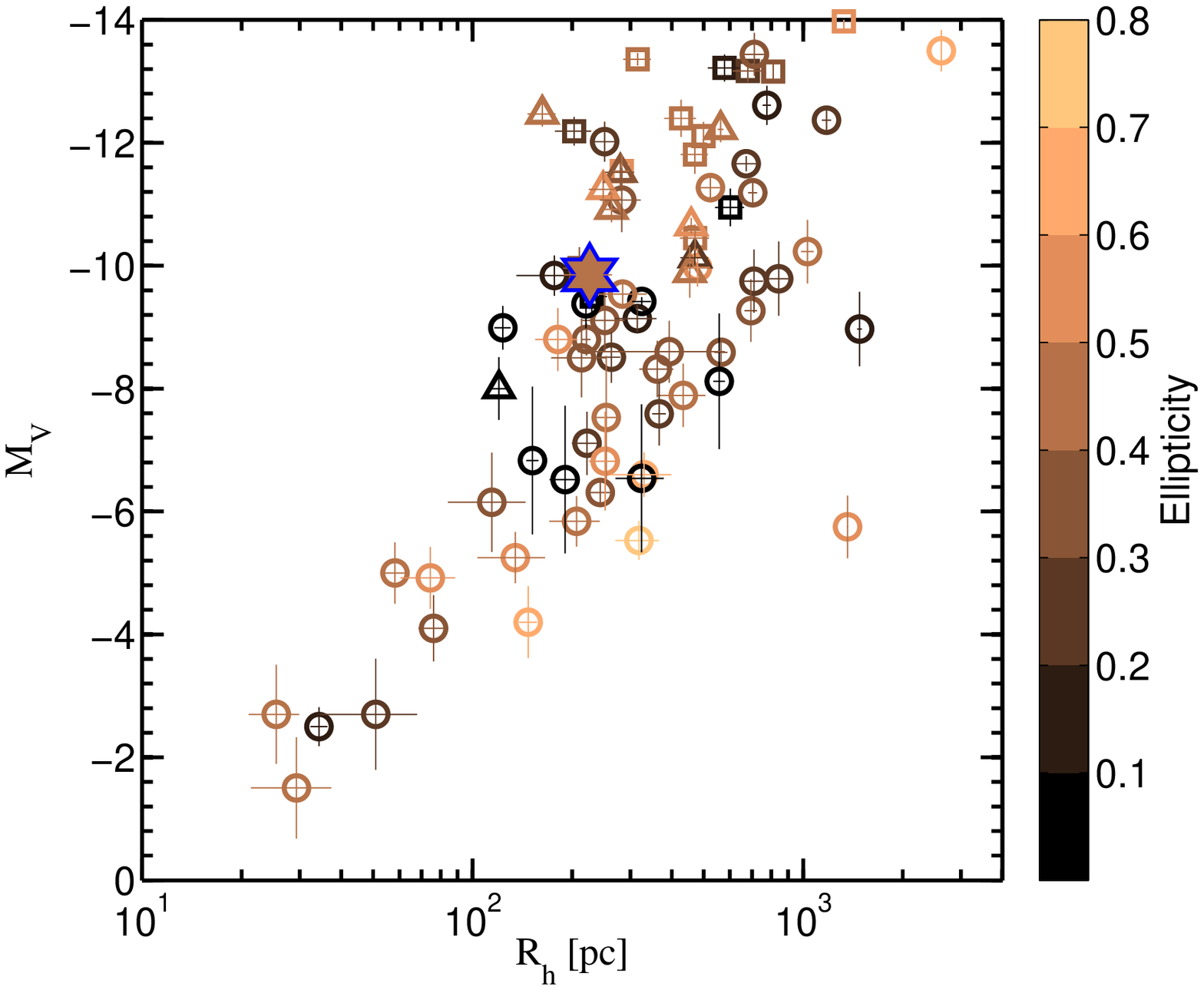}
 \includegraphics[width=88mm, clip]{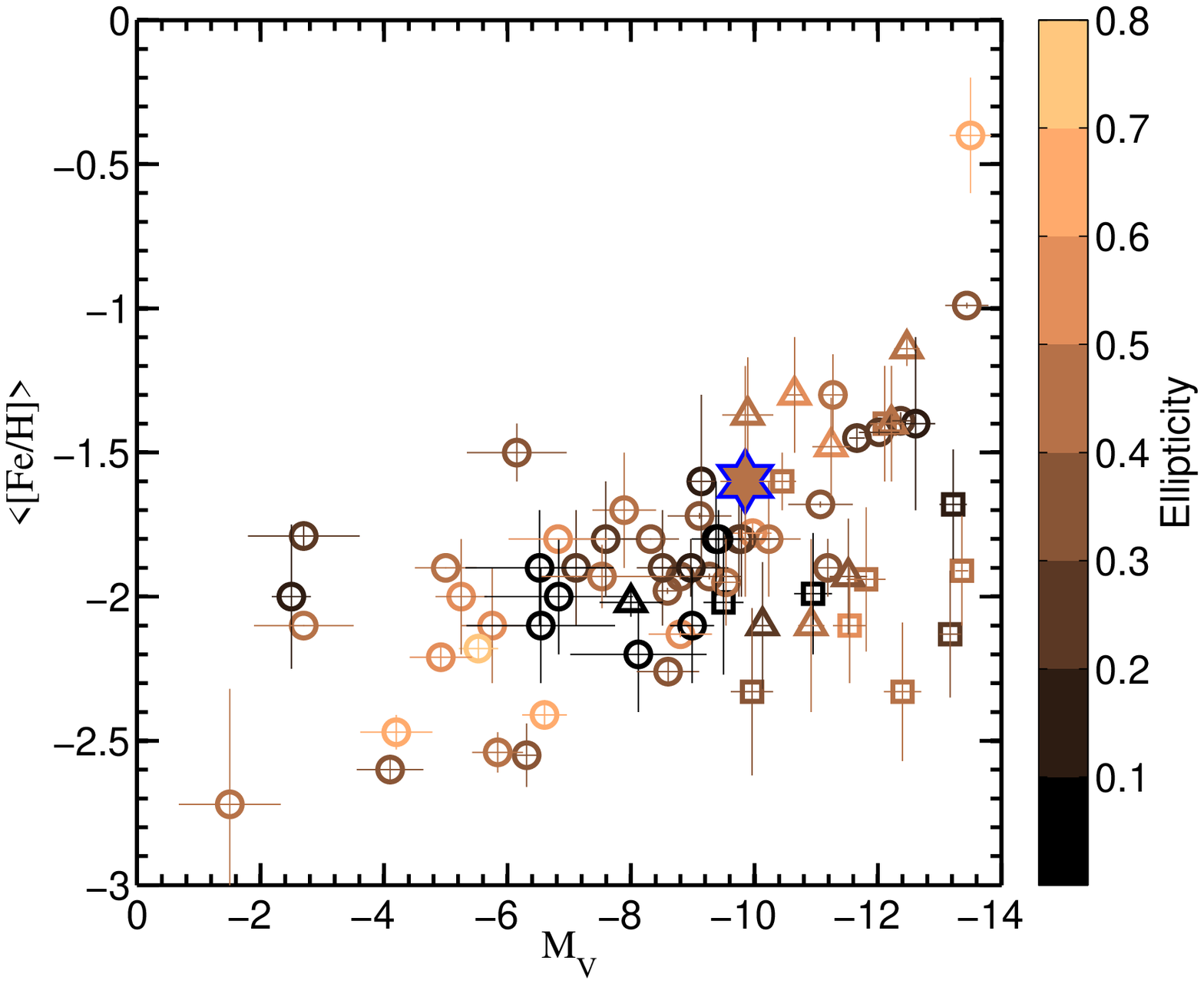}
\caption{Properties of known dwarf galaxies with $M_V > -14$ from data
  compiled in \citet[][see references therein]{McConnachie12}. Left
  panel: Absolute magnitude as a function of size for Local Group and
  nearby dwarfs within 3 Mpc. The color coding indicates the
  ellipticity of such objects; black color is used when there was no
  ellipticity information. dSphs, dIrrs, and transitional type dwarfs
  (or dwarfs whose dSph/dIrr classification is undetermined) are
  represented as circles, squares, and triangles, respectively. GHOSTS
  I is shown as a star. Right panel: stellar metallicity--luminosity
  relation for dwarf galaxies. Symbols are the same as in the left
  panel. The properties of GHOSTS I are similar to those of nearby
  dwarf galaxies at its absolute magnitude.}
\label{fig:dwarfprop}
\end{figure*}

We report the discovery of a faint dwarf galaxy, GHOSTS I, which was
detected by visual inspection of one of the HST/ACS GHOSTS fields in
the halo of M81. While its integrated light is visible, we resolve
individual stars in the dwarf, detecting one magnitude range of its
LF. These stars are most likely very young HeB and AGB stars,
  yielding a distance estimate for GHOSTS I of $\sim 12\pm2$ Mpc. Due
to its distance and small size on the sky (its half-light radius is
$\sim 4\arcsec$), the individual stars of GHOSTS I could have not been
resolved by a telescope other than HST. In Figure~\ref{fig:dwarfprop}
we show the luminosity--size and luminosity--stellar metallicity
relations of Local Group and nearby dwarf galaxies \citep[within 3
  Mpc;][]{McConnachie12}. GHOSTS I's derived properties (see
Table~\ref{table:properties}) are typical of nearby dwarf galaxies at
its absolute magnitude. Due to the relatively large photometric
uncertainties of the star's colors and uncertainties in the
isochrone-based metallicities for low metallicities
\citep[see][Streich et al. 2013]{Monachesi13}, it is difficult to
measure the metallicity of GHOSTS I; we expect it to be low
($[\mathrm{Fe/H}] \lesssim -1.5$) according to the model CMDs that
were used in the previous section to compare with the observed CMD.

GHOSTS I is close on the sky to M81 (NGC 3031) with a radial projected
distance of only 32 kpc from M81's center, assuming M81's
distance. However, M81's distance is 3.7 Mpc (R-S11), whereas we
estimate GHOSTS I's distance to be between 7.9 Mpc and 12.5 Mpc. We
searched for giant (luminous and massive) galaxies and clusters of
galaxies with heliocentric distance $< 16.5$ Mpc and within a
projected radial distance of $25^{\circ}$ from the dwarf's location on
the sky, using the updated galaxy catalog by \citet{Karachentsev13},
the distance estimates for 1791 Galaxies from \citet{Tully08}, and
NED\footnote{\url{http://ned.ipac.caltech.edu/}}. We found that there
are no giant galaxies within 4 Mpc from GHOSTS I. The nearest large
galaxy is NGC 3027, at $\sim 5$ Mpc from the dwarf galaxy, which is
one magnitude fainter than M81.

In general, the majority of dwarfs far away from a large galaxy appear
to be dIrrs, while those closer to their host galaxies tend to be
dSphs. One of the main differences between dIrrs and dSphs is that the
latter lack gas and recent star formation, i.e. they stopped forming
stars for at least a billion years. On the other hand, dIrrs still
retain gas and show evidence of recent star formation, with young
stellar populations of ages $ < 2$ Gyr \citep[see review by][and
  references therein]{Tolstoy09, McConnachie12}. It has also been
noted that dIrrs (within 4 Mpc of the Milky Way) are preferentially
found in lower density environments and have higher luminosities at a
given size than dSphs \citep[][]{Weisz11}. The difference in
luminosity can be observed in the luminosity-size relation shown in
Figure~\ref{fig:dwarfprop}, where squares and circles represent dIrrs
and dSphs, respectively. 

Given their differences in SFHs \citep{Tolstoy09, Weisz11}, the most
direct way to distinguish between dIrrs and dSph is to observe or rule
out young main sequence (MS) and blue and red helium-burning stars
(BHeB and RHeB). GHOSTS I seems to have star forming regions (see red
dashed ellipse in Fig.~\ref{fig:dwarf}), it is morphologically rather
clumpy, i.e. not as smooth as a dSph would be, and from the
  analysis performed in section~\ref{distance}, the most favored
  explanation for their detected stars is that they are young HeB and
  AGB stars. Also, from its SDSS color, $g-r = 0.45\pm 0.11$, it
would be considered a star-forming galaxy according to \citet[][see
  their Figure 3]{Geha12}, who studied the relative number of quenched
versus star-forming dwarf galaxies from SDSS DR8, where quenched
galaxies are defined as those which have no $\textrm{H}\alpha$
emission and a strong 4000 \AA~ break. Furthermore, GHOSTS I is rather
isolated (at least $\sim 4$ Mpc from any giant galaxy), which favors a
dIrr, since dSphs are generally found within 1 Mpc from their host
galaxy. Thus, we tentatively classify it as a dIrr. Deeper HST images,
that could allow us to obtain a CMD detecting at least a 3 magnitude
range of GHOSTS I's LF are required to elucidate its true nature. In
addition, follow-up spectroscopy of this object will allow us to
determine if there are emission lines.

In order to place constraints on the HI content of GHOSTS I, we use
the data cube obtained by \citet{Chynoweth08} using Robert C.  Byrd
Green Bank Telescope (GBT) observations of a $3^{\circ} \times
3^{\circ}$ area centered on the M81 group of galaxies (2003 June 12,
Proposal Code $=$ AGBT03B\_034\_01). Their data cover a very wide velocity
range, from $-605$ to 1970 km s$^{-1}$, excluding the $-85$ to 25 km
s$^{-1}$ velocity window. Given the distance of GHOSTS I, one expects
recession velocities between 700 and 1000 km s$^{-1}$, depending
somewhat on peculiar motions. We find that there is no statistically
significant line emission in that velocity range (smoothed to 15 km
s$^{-1}$, typical for the velocity width of a dwarf galaxy) at the
location of GHOSTS I, placing a 5$\sigma$ upper limit of its HI mass
fraction to $\textrm{M}_{\textrm{HI}}/\textrm{L}_V < 10.4$
M$_\odot$/L$_\odot$.

 If GHOSTS I is indeed a dIrr, it would be, together with Leo P
 \citep{Rhode13} and Leo T \citep{Irwin07}, one of the faintest and
 lowest mass star-forming dwarf galaxies. Studying this object will
 help us understand how these small galaxies retain gas and are able
 to form stars. If, on the contrary, GHOSTS I is a dSph, it would be
 the most distant dSph from any large galaxy ever detected, and
 theoretical models would need to explain how GHOSTS I acquired its
 morphology without being affected by encounters with a giant galaxy.

\acknowledgments AM \& EFB are grateful to Mario Mateo for valuable
discussions on the different possible stellar populations represented
by the dwarf's stars. We thank Katie (Chynoweth) Keating for providing
the full M81 group HI data cubes. We wish to thank the anonymous
  referee for very useful comments and suggestions that helped to
  improve this work. This work was supported by HST grant GO-11613,
provided by NASA through a grant from the Space Telescope Science
Institute, which is operated by the Association of Universities for
Research in Astronomy, Inc., under NASA contract NAS5-26555. The
National Radio Astronomy Observatory is a facility of the National
Science Foundation operated under cooperative agreement by Associated
Universities, Inc. This work has made use of the IAC-STAR synthetic
CMD computation code which is supported and maintained by the computer
division of the Instituto de Astrof\'isica de Canarias, and also of
the NASA/IPAC Extragalactic Database (NED) which is operated by the
Jet Propulsion Laboratory, California Institute of Technology, under
contract with the National Aeronautics and Space Administration.

\emph{Facility:} \facility{HST (ACS)}

\bibliographystyle{apj}
\bibliography{ref}

\begin{thebibliography}{51}
\expandafter\ifx\csname natexlab\endcsname\relax\def\natexlab#1{#1}\fi

\bibitem[{{Aparicio} \& {Gallart}(2004)}]{Aparicio_gallart04}
{Aparicio}, A., \& {Gallart}, C. 2004, \aj, 128, 1465

\bibitem[{{Bell} {et~al.}(2008){Bell}, {Zucker}, {Belokurov}, {Sharma},
  {Johnston}, {Bullock}, {Hogg}, {Jahnke}, {de Jong}, {Beers}, {Evans},
  {Grebel}, {Ivezi{\'c}}, {Koposov}, {Rix}, {Schneider}, {Steinmetz}, \&
  {Zolotov}}]{Bell08}
{Bell}, E.~F., {et~al.} 2008, \apj, 680, 295

\bibitem[{{Bellazzini} {et~al.}(2001){Bellazzini}, {Ferraro}, \&
  {Pancino}}]{Bellazzini01}
{Bellazzini}, M., {Ferraro}, F.~R., \& {Pancino}, E. 2001, \apj, 556, 635

\bibitem[{{Belokurov} {et~al.}(2007){Belokurov}, {Zucker}, {Evans}, {Kleyna},
  {Koposov}, {Hodgkin}, {Irwin}, {Gilmore}, {Wilkinson}, {Fellhauer},
  {Bramich}, {Hewett}, {Vidrih}, {De Jong}, {Smith}, {Rix}, {Bell}, {Wyse},
  {Newberg}, {Mayeur}, {Yanny}, {Rockosi}, {Gnedin}, {Schneider}, {Beers},
  {Barentine}, {Brewington}, {Brinkmann}, {Harvanek}, {Kleinman}, {Krzesinski},
  {Long}, {Nitta}, \& {Snedden}}]{Belokurov07}
{Belokurov}, V., {et~al.} 2007, \apj, 654, 897

\bibitem[{{Benson} {et~al.}(2002){Benson}, {Frenk}, {Lacey}, {Baugh}, \&
  {Cole}}]{Benson02}
{Benson}, A.~J., {Frenk}, C.~S., {Lacey}, C.~G., {Baugh}, C.~M., \& {Cole}, S.
  2002, \mnras, 333, 177

\bibitem[{{Boylan-Kolchin} {et~al.}(2012){Boylan-Kolchin}, {Bullock}, \&
  {Kaplinghat}}]{Boylan-Kolchin12}
{Boylan-Kolchin}, M., {Bullock}, J.~S., \& {Kaplinghat}, M. 2012, \mnras, 422,
  1203

\bibitem[{{Brooks} {et~al.}(2013){Brooks}, {Kuhlen}, {Zolotov}, \&
  {Hooper}}]{Brooks13}
{Brooks}, A.~M., {Kuhlen}, M., {Zolotov}, A., \& {Hooper}, D. 2013, \apj, 765,
  22

\bibitem[{{Bullock} \& {Johnston}(2005)}]{BJ05}
{Bullock}, J.~S., \& {Johnston}, K.~V. 2005, \apj, 635, 931 (BJ05)

\bibitem[{{Bullock} {et~al.}(2000){Bullock}, {Kravtsov}, \&
  {Weinberg}}]{Bullock00}
{Bullock}, J.~S., {Kravtsov}, A.~V., \& {Weinberg}, D.~H. 2000, \apj, 539, 517

\bibitem[{{Bullock} {et~al.}(2001){Bullock}, {Kravtsov}, \&
  {Weinberg}}]{Bullock01}
---. 2001, \apj, 548, 33

\bibitem[{{Chiboucas} {et~al.}(2013){Chiboucas}, {Jacobs}, {Tully}, \&
  {Karachentsev}}]{Chiboucas13}
{Chiboucas}, K., {Jacobs}, B.~A., {Tully}, R.~B., \& {Karachentsev}, I.~D.
  2013, \aj, 146, 126

\bibitem[{{Chiboucas} {et~al.}(2009){Chiboucas}, {Karachentsev}, \&
  {Tully}}]{Chiboucas09}
{Chiboucas}, K., {Karachentsev}, I.~D., \& {Tully}, R.~B. 2009, \aj, 137, 3009

\bibitem[{{Chynoweth} {et~al.}(2008){Chynoweth}, {Langston}, {Yun}, {Lockman},
  {Rubin}, \& {Scoles}}]{Chynoweth08}
{Chynoweth}, K.~M., {Langston}, G.~I., {Yun}, M.~S., {Lockman}, F.~J., {Rubin},
  K.~H.~R., \& {Scoles}, S.~A. 2008, \aj, 135, 1983

\bibitem[{{Cooper} {et~al.}(2010){Cooper}, {Cole}, {Frenk}, {White}, {Helly},
  {Benson}, {De Lucia}, {Helmi}, {Jenkins}, {Navarro}, {Springel}, \&
  {Wang}}]{Cooper10}
{Cooper}, A.~P., {et~al.} 2010, \mnras, 406, 744

\bibitem[{{Crnojevi{\'c}} {et~al.}(2012){Crnojevi{\'c}}, {Grebel}, \&
  {Cole}}]{Crnojevic12}
{Crnojevi{\'c}}, D., {Grebel}, E.~K., \& {Cole}, A.~A. 2012, \aap, 541, A131

\bibitem[{{Dalcanton} {et~al.}(2009){Dalcanton}, {Williams}, {Seth}, {Dolphin},
  {Holtzman}, {Rosema}, {Skillman}, {Cole}, {Girardi}, {Gogarten},
  {Karachentsev}, {Olsen}, {Weisz}, {Christensen}, {Freeman}, {Gilbert},
  {Gallart}, {Harris}, {Hodge}, {de Jong}, {Karachentseva}, {Mateo}, {Stetson},
  {Tavarez}, {Zaritsky}, {Governato}, \& {Quinn}}]{Dalcanton09}
{Dalcanton}, J.~J., {et~al.} 2009, \apjs, 183, 67

\bibitem[{{Dekel} \& {Silk}(1986)}]{DekelSilk86}
{Dekel}, A., \& {Silk}, J. 1986, \apj, 303, 39

\bibitem[{{Dolphin}(2000)}]{Dolphin00}
{Dolphin}, A.~E. 2000, \pasp, 112, 1383

\bibitem[{{Geha} {et~al.}(2012){Geha}, {Blanton}, {Yan}, \& {Tinker}}]{Geha12}
{Geha}, M., {Blanton}, M.~R., {Yan}, R., \& {Tinker}, J.~L. 2012, \apj, 757, 85

\bibitem[{{G{\'o}mez} {et~al.}(2012){G{\'o}mez}, {Coleman-Smith}, {O'Shea},
  {Tumlinson}, \& {Wolpert}}]{Gomez12}
{G{\'o}mez}, F.~A., {Coleman-Smith}, C.~E., {O'Shea}, B.~W., {Tumlinson}, J.,
  \& {Wolpert}, R.~L. 2012, \apj, 760, 112

\bibitem[{{Irwin} {et~al.}(2007){Irwin}, {Belokurov}, {Evans}, {Ryan-Weber},
  {de Jong}, {Koposov}, {Zucker}, {Hodgkin}, {Gilmore}, {Prema}, {Hebb},
  {Begum}, {Fellhauer}, {Hewett}, {Kennicutt}, {Wilkinson}, {Bramich},
  {Vidrih}, {Rix}, {Beers}, {Barentine}, {Brewington}, {Harvanek},
  {Krzesinski}, {Long}, {Nitta}, \& {Snedden}}]{Irwin07}
{Irwin}, M.~J., {et~al.} 2007, \apjl, 656, L13

\bibitem[{{Jang} {et~al.}(2012){Jang}, {Lim}, {Park}, \& {Lee}}]{Jang12}
{Jang}, I.~S., {Lim}, S., {Park}, H.~S., \& {Lee}, M.~G. 2012, \apjl, 751, L19

\bibitem[{{Karachentsev} {et~al.}(2013){Karachentsev}, {Makarov}, \&
  {Kaisina}}]{Karachentsev13}
{Karachentsev}, I.~D., {Makarov}, D.~I., \& {Kaisina}, E.~I. 2013, \aj, 145,
  101

\bibitem[{{Karachentsev} {et~al.}(2002){Karachentsev}, {Sharina}, {Dolphin},
  {Grebel}, {Geisler}, {Guhathakurta}, {Hodge}, {Karachentseva}, {Sarajedini},
  \& {Seitzer}}]{Karachentsev02}
{Karachentsev}, I.~D., {et~al.} 2002, \aap, 385, 21

\bibitem[{{Kazantzidis} {et~al.}(2013){Kazantzidis}, {{\L}okas}, \&
  {Mayer}}]{Kazantzidis13}
{Kazantzidis}, S., {{\L}okas}, E.~L., \& {Mayer}, L. 2013, \apjl, 764, L29

\bibitem[{{Klypin} {et~al.}(1999){Klypin}, {Kravtsov}, {Valenzuela}, \&
  {Prada}}]{Klypin99}
{Klypin}, A., {Kravtsov}, A.~V., {Valenzuela}, O., \& {Prada}, F. 1999, \apj,
  522, 82

\bibitem[{{Koposov} {et~al.}(2008){Koposov}, {Belokurov}, {Evans}, {Hewett},
  {Irwin}, {Gilmore}, {Zucker}, {Rix}, {Fellhauer}, {Bell}, \&
  {Glushkova}}]{Koposov08}
{Koposov}, S., {et~al.} 2008, \apj, 686, 279

\bibitem[{{Koposov} {et~al.}(2009){Koposov}, {Yoo}, {Rix}, {Weinberg},
  {Macci{\`o}}, \& {Escud{\'e}}}]{Koposov09}
{Koposov}, S.~E., {Yoo}, J., {Rix}, H.-W., {Weinberg}, D.~H., {Macci{\`o}},
  A.~V., \& {Escud{\'e}}, J.~M. 2009, \apj, 696, 2179

\bibitem[{{Makarov} {et~al.}(2012){Makarov}, {Makarova}, {Sharina}, {Uklein},
  {Tikhonov}, {Guhathakurta}, {Kirby}, \& {Terekhova}}]{Makarov12}
{Makarov}, D., {Makarova}, L., {Sharina}, M., {Uklein}, R., {Tikhonov}, A.,
  {Guhathakurta}, P., {Kirby}, E., \& {Terekhova}, N. 2012, \mnras, 425, 709

\bibitem[{{Martin} {et~al.}(2009){Martin}, {McConnachie}, {Irwin}, {Widrow},
  {Ferguson}, {Ibata}, {Dubinski}, {Babul}, {Chapman}, {Fardal}, {Lewis},
  {Navarro}, \& {Rich}}]{Martin09}
{Martin}, N.~F., {et~al.} 2009, \apj, 705, 758

\bibitem[{{Mayer} {et~al.}(2001){Mayer}, {Governato}, {Colpi}, {Moore},
  {Quinn}, {Wadsley}, {Stadel}, \& {Lake}}]{Mayer01}
{Mayer}, L., {Governato}, F., {Colpi}, M., {Moore}, B., {Quinn}, T., {Wadsley},
  J., {Stadel}, J., \& {Lake}, G. 2001, \apj, 559, 754

\bibitem[{{Mayer} {et~al.}(2006){Mayer}, {Mastropietro}, {Wadsley}, {Stadel},
  \& {Moore}}]{Mayer06}
{Mayer}, L., {Mastropietro}, C., {Wadsley}, J., {Stadel}, J., \& {Moore}, B.
  2006, \mnras, 369, 1021

\bibitem[{{McConnachie}(2012)}]{McConnachie12}
{McConnachie}, A.~W. 2012, \aj, 144, 4

\bibitem[{{Monachesi} {et~al.}(2012){Monachesi}, {Trager}, {Lauer}, {Hidalgo},
  {Freedman}, {Dressler}, {Grillmair}, \& {Mighell}}]{Monachesi12}
{Monachesi}, A., {Trager}, S.~C., {Lauer}, T.~R., {Hidalgo}, S.~L., {Freedman},
  W., {Dressler}, A., {Grillmair}, C., \& {Mighell}, K.~J. 2012, \apj, 745, 97

\bibitem[{{Monachesi} {et~al.}(2013){Monachesi}, {Bell}, {Radburn-Smith},
  {Vlaji{\'c}}, {de Jong}, {Bailin}, {Dalcanton}, {Holwerda}, \&
  {Streich}}]{Monachesi13}
{Monachesi}, A., {et~al.} 2013, \apj, 766, 106

\bibitem[{{Moore} {et~al.}(1999){Moore}, {Ghigna}, {Governato}, {Lake},
  {Quinn}, {Stadel}, \& {Tozzi}}]{Moore99}
{Moore}, B., {Ghigna}, S., {Governato}, F., {Lake}, G., {Quinn}, T., {Stadel},
  J., \& {Tozzi}, P. 1999, \apjl, 524, L19

\bibitem[{{Pasquali} {et~al.}(2005){Pasquali}, {Larsen}, {Ferreras}, {Gnedin},
  {Malhotra}, {Rhoads}, {Pirzkal}, \& {Walsh}}]{Pasquali05}
{Pasquali}, A., {Larsen}, S., {Ferreras}, I., {Gnedin}, O.~Y., {Malhotra}, S.,
  {Rhoads}, J.~E., {Pirzkal}, N., \& {Walsh}, J.~R. 2005, \aj, 129, 148

\bibitem[{{Radburn-Smith} {et~al.}(2011){Radburn-Smith}, {de Jong}, {Seth},
  {Bailin}, {Bell}, {Brown}, {Bullock}, {Courteau}, {Dalcanton}, {Ferguson},
  {Goudfrooij}, {Holfeltz}, {Holwerda}, {Purcell}, {Sick}, {Streich}, {Vlajic},
  \& {Zucker}}]{RS11}
{Radburn-Smith}, D.~J., {et~al.} 2011, \apjs, 195, 18 (RS11)

\bibitem[{{Rhode} {et~al.}(2013){Rhode}, {Salzer}, {Haurberg}, {Van Sistine},
  {Young}, {Haynes}, {Giovanelli}, {Cannon}, {Skillman}, {McQuinn}, \&
  {Adams}}]{Rhode13}
{Rhode}, K.~L., {et~al.} 2013, \aj, 145, 149

\bibitem[{{Richardson} {et~al.}(2011){Richardson}, {Irwin}, {McConnachie},
  {Martin}, {Dotter}, {Ferguson}, {Ibata}, {Chapman}, {Lewis}, {Tanvir}, \&
  {Rich}}]{Richardson11}
{Richardson}, J.~C., {et~al.} 2011, \apj, 732, 76

\bibitem[{{Schlafly} \& {Finkbeiner}(2011)}]{Schlafly11}
{Schlafly}, E.~F., \& {Finkbeiner}, D.~P. 2011, \apj, 737, 103

\bibitem[{{Schlegel} {et~al.}(1998){Schlegel}, {Finkbeiner}, \&
  {Davis}}]{Schlegel98}
{Schlegel}, D.~J., {Finkbeiner}, D.~P., \& {Davis}, M. 1998, \apj, 500, 525

\bibitem[{{Sirianni} {et~al.}(2005){Sirianni}, {Jee}, {Ben{\'{\i}}tez},
  {Blakeslee}, {Martel}, {Meurer}, {Clampin}, {De Marchi}, {Ford}, {Gilliland},
  {Hartig}, {Illingworth}, {Mack}, \& {McCann}}]{Sirianni05}
{Sirianni}, M., {et~al.} 2005, \pasp, 117, 1049

\bibitem[{{Slater} \& {Bell}(2013)}]{Slater_bell13}
{Slater}, C.~T., \& {Bell}, E.~F. 2013, ArXiv e-prints

\bibitem[{{Slater} {et~al.}(2011){Slater}, {Bell}, \& {Martin}}]{Slater11}
{Slater}, C.~T., {Bell}, E.~F., \& {Martin}, N.~F. 2011, \apjl, 742, L14

\bibitem[{{Teyssier} {et~al.}(2012){Teyssier}, {Johnston}, \&
  {Kuhlen}}]{Teyssier12}
{Teyssier}, M., {Johnston}, K.~V., \& {Kuhlen}, M. 2012, \mnras, 426, 1808

\bibitem[{{Tolstoy} {et~al.}(2009){Tolstoy}, {Hill}, \& {Tosi}}]{Tolstoy09}
{Tolstoy}, E., {Hill}, V., \& {Tosi}, M. 2009, \araa, 47, 371

\bibitem[{{Tully} {et~al.}(2008){Tully}, {Shaya}, {Karachentsev}, {Courtois},
  {Kocevski}, {Rizzi}, \& {Peel}}]{Tully08}
{Tully}, R.~B., {Shaya}, E.~J., {Karachentsev}, I.~D., {Courtois}, H.~M.,
  {Kocevski}, D.~D., {Rizzi}, L., \& {Peel}, A. 2008, \apj, 676, 184

\bibitem[{{Walsh} {et~al.}(2009){Walsh}, {Willman}, \& {Jerjen}}]{Walsh09}
{Walsh}, S.~M., {Willman}, B., \& {Jerjen}, H. 2009, \aj, 137, 450

\bibitem[{{Weisz} {et~al.}(2011){Weisz}, {Dalcanton}, {Williams}, {Gilbert},
  {Skillman}, {Seth}, {Dolphin}, {McQuinn}, {Gogarten}, {Holtzman}, {Rosema},
  {Cole}, {Karachentsev}, \& {Zaritsky}}]{Weisz11}
{Weisz}, D.~R., {et~al.} 2011, \apj, 739, 5

\bibitem[{{Willman} {et~al.}(2005){Willman}, {Dalcanton}, {Martinez-Delgado},
  {West}, {Blanton}, {Hogg}, {Barentine}, {Brewington}, {Harvanek}, {Kleinman},
  {Krzesinski}, {Long}, {Neilsen}, {Nitta}, \& {Snedden}}]{Willman05}
{Willman}, B., {et~al.} 2005, \apjl, 626, L85

\end{thebibliography}

\end{document}